\documentclass{azh}

\usepackage{graphicx}
\usepackage{natbib}
\usepackage{lscape}

\def\sun{\hbox{$\odot$}}
\def\degr{\hbox{$^\circ$}}
\def\arcmin{\hbox{$^\prime$}}
\def\arcsec{\hbox{$^{\prime\prime}$}}

\begin{document} 

\title{Parameters of Star Formation Regions in Galaxies \\ 
       NGC~3963 and NGC~7292}
       
\author{A.~S.~Gusev,$^1$
        F.~Kh.~Sakhibov,$^2$
        A.~V.~Moiseev,$^{3,1}$
        V.~S.~Kostiuk,$^1$
        and D.~V.~Oparin$^3$}

\institute{$^1$ Sternberg Astronomical Institute, Moscow State University, Moscow, Russia \\
           $^2$ University of Applied Sciences, Technische Hochschule Mittelhessen, Friedberg, 
                Germany \\
           $^3$ Special Astrophysical Observatory, Russian Academy of Sciences, Nizhnii Arkhyz, 
           369167 Russia}

\date{Received March 6, 2024; revised May 2, 2024; accepted June 4, 2024}
\offprints{Alexander~S.~Gusev, \email{gusev@sai.msu.ru}}

\titlerunning{Parameters of Star Formation Regions}
\authorrunning{Gusev et al.}

\abstract{Results of a study of physical parameters of stellar population in star 
formation regions in galaxies with signs of peculiarity NGC~3963 and NGC~7292 are 
presented. The study was carried out based on the analysis of photometric ($UBVRI$ 
bands), H$\alpha$ and spectroscopic data obtained by the authors, using 
evolutionary models of stellar population. Among 157 star formation regions identified 
in galaxies, the young stellar population mass estimates were obtained for 16 of them 
and the age estimates were obtained for 15. The age of star formation regions clearly 
correlates with the presence of emission in the H$\alpha$ line: HII~regions in the 
galaxies are younger than 6-8~Myr, and the regions without gas emission are older. 
The studied objects are included in the version~3 of our catalogue of photometric, 
physical and chemical parameters of star formation regions, which includes 1667 objects 
in 21 galaxies. Key aspects of the technique used to estimate the physical parameters 
and different relations between observational and physical parameters of the young 
stellar population in star formation regions are discussed. \\

{\bf Keywords:} spiral galaxies, peculiar galaxies, star formation, HII~regions, 
stellar population \\

{\bf DOI:} 10.1134/S1063772924700616 \\
}

\maketitle

\section{INTRODUCTION}

The processes of modern star formation, observed over a wide range of 
wavelengths (from ultraviolet to infrared) in most spiral and some 
lenticular galaxies, are among the most visible indicators of galaxy 
evolution in the present epoch. Stars form in groups within the densest 
and coldest parts of giant molecular hydrogen clouds. The largest groups of 
newborn stars, known as stellar complexes, can reach sizes of 600--700~pc 
\citep{efremov1989,elmegreen1994,efremov1995,elmegreen1996,efremov1998,
odekon2008,elmegreen2009,marcos2009,zwart2010}. These complexes consist of 
smaller groupings such as OB associations, star clusters, and star aggregates 
\citep{efremov1987,ivanov1991}.

The main indicator of star formation in the optical range is the emission 
from HII regions in the Balmer hydrogen lines, primarily in the H$\alpha$ 
line. The ionization of hydrogen is caused by the most massive stars. 
However, star formation regions (SFRs) younger than 1--2~Myr are not 
observed in optical wavelengths due to significant extinction by the dust 
cocoon surrounding the star--gas grouping 
\citep{whitmore2011,hollyhead2015,kim2021}. The lifespan of an HII shell 
does not exceed 8-10~Myr.

Data on the physical and chemical parameters of stars and gas in SFRs, 
such as size, metallicity, mass, and age, are crucial for understanding 
modern evolutionary processes in disk galaxies. Particularly challenging 
in unresolved groupings are the correct accounting for extinction in SFRs, 
the separation of the contributions of stellar and gas emissions in broad 
photometric bands, and the accounting for the effect of discreteness of 
the initial mass function (IMF) for low-mass star clusters and associations 
\citep{piskunov2011,cervino2013,gusev2016,gusev2018,gusev2023}. The 
research method used will be described in detail in 
Section~\ref{sect:method}.

\begin{figure*}
\centerline{\includegraphics[width=17.8cm]{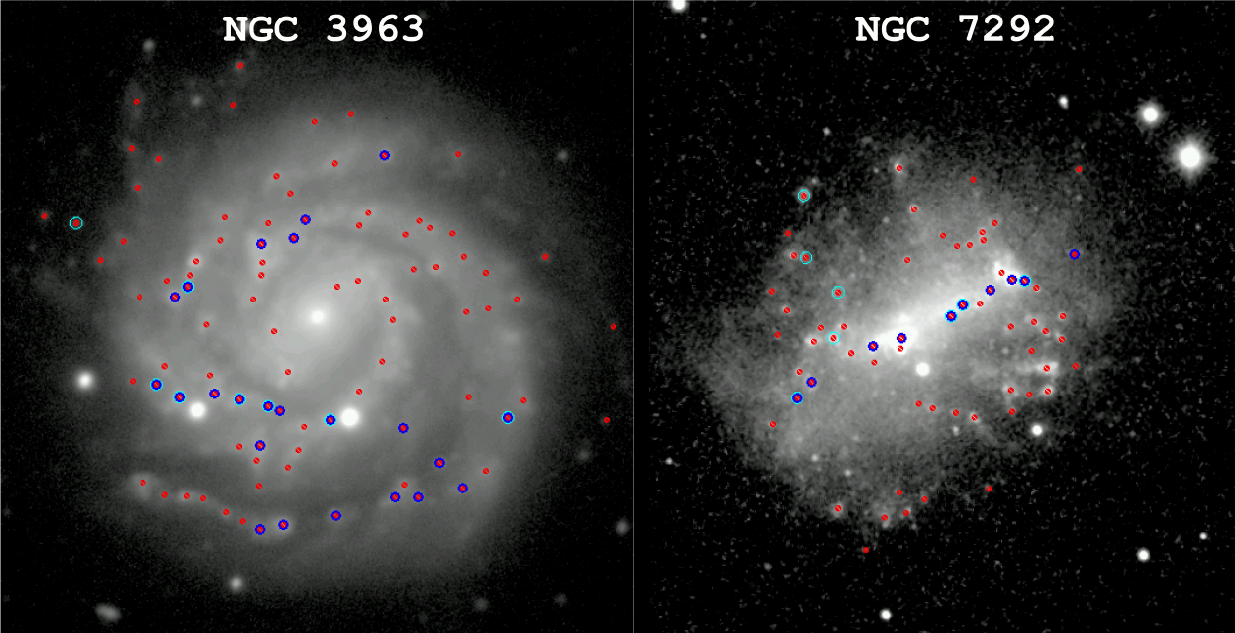}}
\caption{Images of the galaxies NGC~3963 (left) and NGC~7292 (right) in the band 
on a logarithmic intensity scale. Red small circles indicate identified SFRs 
studied photometrically, blue circles denote SFRs studied spectroscopically in 
\citet{gusev2021}, and large cyan circles represent SFRs with obtained mass (age) 
estimates in this study. The image size for NGC~3963 is 
$170.5\arcsec\times170.5\arcsec$, corresponding to a linear size of 40.67~kpc, 
and the image size for NGC~7292 is $146.7\arcsec\times146.7\arcsec$ (4.85~kpc). 
North is up, east is to the left. The centers of the images correspond to the 
centers of the galaxies.
\label{fig:map}}
\end{figure*}

In recent years, several large international projects such as 
LEGUS \citep{calzetti2015}, PHANGS-MUSE \citep{emsellem2022}, and 
PHANGS-{\it HST} \citep{lee2022} facilitated the study of tens of thousands 
of young star clusters, OB associations, and HII~regions in many galaxies 
\citep{adamo2017,whitmore2021,thilker2022,larson2023,groves2023}. Generally, 
HII~regions and young stellar groupings are considered separately within 
these projects. In particular, the age and mass of young stellar groupings 
and extinction in the star formation region are determined by comparing the 
spectral energy distribution with theoretical evolutionary models 
\citep[see, e.g.,][]{turner2021} without using extinction data from HII~regions 
obtained from spectroscopic observations. In recent years, several studies have 
jointly considered the properties of stellar populations and gas in HII~regions. 
However, even in these studies, the age and mass of the stellar population are 
determined without extinction data from spectral observations. For example, 
in \citet{scheuermann2023}, the authors studied the properties of stellar 
populations in HII~regions by comparing data from the PHANGS-MUSE and 
PHANGS-{\it HST} catalogs but did not use Balmer extinction data to estimate the 
ages of young stars. Unlike those studies, we investigate the properties of 
stellar populations in HII~regions using both photometric and spectral data (see 
Section~\ref{sect:method} for more details).

The main goal of the study is to estimate the physical parameters (mass and age) 
of the stellar population in SFRs based on comprehensive photometric, 
spectrophotometric, and spectroscopic observations of galaxies. Previously, 
in our series of works \citep[see][and references therein]{gusev2023}, we 
reviewed and analyzed a sample of 1510 SFRs in 19 galaxies based on photometric 
data in the $UBVRI$ and the H$\alpha$ line, as well as spectroscopic results of 
the associated HII~regions. The catalog of studied regions is available in 
electronic form.\footnote{http://lnfm1.sai.msu.ru/$\sim$gusev/sfr\_cat.html} 
In this study, we extend our sample to SFRs in two more galaxies: NGC~3963 
and NGC~7292.

Both galaxies studied by us using spectroscopy in \citet{gusev2021} are 
interesting due to their various peculiarities. The SAB(rs)bc-type galaxy 
NGC~3963 has an almost perfectly symmetrical shape (Fig.~\ref{fig:map}). 
However, an excess of oxygen and nitrogen is observed in the HII~regions 
in the outer part of the galaxy's southern spiral arm, possibly explained by 
the inflow of metal-enriched gas into the southwest part of NGC~3963 
\citep{gusev2021}. The shape of the southern spiral arm deviates from a classic 
logarithmic spiral (Fig.~\ref{fig:map}), and HI observational data analysis 
results indicate tidal distortions in NGC~3963 and the galaxy NGC~3958, located 
110~kpc southwest of it \citep{moorsel1983}. The Magellanic-type galaxy 
NGC~7292, with a bright asymmetric bar (Fig.~\ref{fig:map}), has rather complex 
kinematics: radial movements related to the bar play a significant role, and some 
non-circular motions at the southeast end of the bar -- the brightest HII~region 
-- may be associated with the aftermath of a merger with a companion 
\citep{gusev2023b}.

\begin{table}
\begin{center}
\caption{Main characteristics of the galaxies}
\label{tab:gen}
\begin{tabular}{l|c|c}
\hline\hline
Parameter & NGC~3963 & NGC~7292 \\
\hline
Coordinates & $11^h54^m58.7^s$ & $22^h28^m25.3^s$ \\
of the center: $\alpha$, $\delta$ (J2000) & $+58\degr 29\arcmin 37.1\arcsec$ & 
$+30\degr 17\arcmin 35.3\arcsec$ \\
Type & SAB(rs)bc & IBm \\
$m(B)$, mag & $12.60\pm0.07$ & $13.06\pm0.06$ \\
$M(B)_0^i$, mag & $-21.1\pm0.3$ & $-16.7\pm1.0$ \\
$i$, degr & 29 & 29 \\
PA, degr & 96 & 73 \\
$d$, Mpc & 49.2 & 6.82 \\
$D_{25}$, arcmin & 2.57 & 1.91 \\
$D_{25}$, kpc & 36.8 & 3.78 \\
$A(B)_{\rm G}$, mag & 0.083 & 0.223 \\
$A(B)_{\rm in}$, mag & 0.09 & 0.14 \\
\hline
\end{tabular}
\end{center}
\end{table}

Basic information about the galaxies -- coordinates of the center, 
morphological type, apparent magnitude $m(B)$, corrected for galactic 
extinction and disk inclination extinction, absolute magnitude 
$M(B)^i_0$, inclination $i$ and positional angle PA of the disk, 
distance $d$, diameter at the isophote $25^m$ in the $B$ band considering 
galactic extinction and extinction caused by the disc inclination 
($D_{25}$), galactic extinction $A(B)_{\rm G}$, and extinction caused 
by the inclination of the galactic disk $A(B)_{\rm in}$ -- are provided 
in Table~\ref{tab:gen}. The galaxy data were taken from the 
NED\footnote{http://ned.ipac.caltech.edu/} and 
HyperLEDA\footnote{http://leda.univ-lyon1.fr/} open databases. 
Exceptions are the data for the center coordinates, inclination, and 
positional angle of the NGC~7292 galaxy, for which we used values from 
\citet{gusev2023b}.

\section{OBSERVATIONS AND DATA REDUCTION}

Spectral observations with a long slit conducted in 2020 on the 
2.5-m telescope at the Caucasian Mountain Observatory (CMO) of 
the Sternberg Astronomical Institute of Moscow State University 
(SAI MSU) using the transient double-beam spectrograph (TDS) 
\citep{potanin2020}, are detailed in \citet{gusev2021}. In total, 
data for 23 HII~regions in NGC~3963 (including several objects 
previously observed within the SDSS 
project\footnote{http://skyserver.sdss.org/dr16}) and 9 in 
NGC~7292 were obtained and analyzed.

Photometric observations of NGC~3963 in $UBVRI$ bands were conducted 
in 2020 on the 2.5-m telescope at the CMO SAI MSU at the Cassegrain 
focus f/8 (see the observation log in Table~\ref{tab:obs}). The NBI 
CCD camera, equipped with two detectors of $2048\times4102$~px, 
provided an image field of $10.5\arcmin\times10.5\arcmin$ at a scale 
of $0.155\arcsec$/px. Observations of NGC~7292 were conducted in 2005 
on the 1.5-m telescope at Maidanak Observatory of the Astronomical 
Institute of the Academy of Sciences of Uzbekistan (Uzbekistan) using 
the SITe-2000 CCD camera at the 1:8 focus. The matrix size was 
$2000\times800$~px, providing a field of view of 
$8.9\arcmin\times3.6\arcmin$ at a scale of $0.267\arcsec$/px 
\citep{artamonov2010}. Further processing followed the standard procedure 
using the ESO-MIDAS image processing system \citep[see, e.g.][]{bruevich2010}. 
To construct color equations and account for atmospheric extinction, 
observations of standard stars from the Landolt fields PG~0231+051, 
PG~2331+055, J125239+444615, GD~279, and GD~300 
\citep{landolt1992,landolt2013,landolt2016} were used, obtained on the 
same nights in the corresponding filters. The color transformation equations 
from the instrumental photometric system to the standard Johnson--Cousins 
$UBVRI$ system are detailed in \citet{artamonov2010} for the Maidanak 
Observatory telescope and in \citet{gusev2018} for the CMO.

\begin{table}
\begin{center}
\caption{Observation log}
\label{tab:obs}
\begin{tabular}{l|c|c|c|c}
\hline\hline
Galaxy & Dates & Band & Expo- & $\epsilon$*  \\
       &       &      & sure, s   & \\
\hline
NGC  & 13/14.04.2020, & $U$ & 3000 & 2.0$\arcsec$ \\
3963 & 14/15.12.2020  & $B$ & 1500 & 2.2 \\
     &                & $V$ &  750 & 2.1 \\
     &                & $R$ &  450 & 1.9 \\
     &                & $I$ &  300 & 1.8 \\
     & 07/08.11.2021  & H$\alpha$ & 2100 & 1.8 \\
     &                & [NII] & 1200 & 1.7 \\
     &                & H$\alpha_{\rm cont}$ & 1800 & 1.9 \\
NGC  & 25/26.10.2005  & $U$ & 1200 & 1.4 \\
7292 &                & $B$ &  600 & 0.9 \\
     &                & $V$ &  480 & 1.0 \\
     &                & $R$ &  360 & 1.0 \\
     &                & $I$ &  240 & 0.8 \\
     & 13/14.12.2020, & H$\alpha$+[NII] & 1050 & 0.9 \\
     & 14/15.12.2020  & H$\alpha_{\rm cont}$ & 1200 & 0.9 \\
\hline
\end{tabular}
\end{center}
*$\epsilon$ is the image seeing.
\end{table}

The galaxy NGC~3963 is quite distant from us (see Table~\ref{tab:gen}), 
and its H$\alpha$ emission lies outside the narrow-band H$\alpha$+[NII] 
filters available in the NBI camera set, centered at a wavelength of 
$\lambda=6563$\AA, with a typical bandwidth $\Delta\lambda\sim60-80$\AA. 
Therefore, its observations in the H$\alpha$ and [NII]$\lambda$6584 lines 
were conducted in 2021 on the 2.5-m telescope at CMO in the Nasmyth focus 
using the MaNGaL narrow emission line mapper -- a photometer with a tunable 
filter based on a low-order Fabry-Perot interferometer with an instrumental 
$FWHM=13$\AA \, \citep{moiseev2020}. The detector of the instrument is 
the Andor iKon-M934 CCD camera with $1024\times1024$~px, providing images 
of $5.6\arcmin\times5.6\arcmin$ at a scale of $0.33\arcsec$/px on this 
telescope. During the observations, images were sequentially obtained at 
wavelengths corresponding to the H$\alpha$, [NII]$\lambda$6584 emission 
lines and the continuum at $40-50$~\AA \, both sides from the H$\alpha$ 
line, considering the systemic velocity of the galaxy. The duration of 
individual exposures was 300~s, and only frames with the best image 
quality were used in the further analysis. The total exposure times are 
listed in Table~\ref{tab:obs}. The observations were processed according 
to the algorithms described in \citet{moiseev2020}, and the absolute flux 
calibration was performed using images of the spectrophotometric standard 
AGK+81$\degr$266, obtained on the same night just before the galaxy 
observations, at a similar air mass.

Observations in the H$\alpha$+[NII] filter for the galaxy NGC~7292 were 
conducted in December 2020 on the 2.5-m telescope at CMO using the same NBI 
camera as in the photometric observations of NGC~3963. The standard 
narrowband H$\alpha$+[NII] filter at the observatory has a peak 
transmission at $\lambda_{\rm cent}=6558$\AA \, and a half-width 
$\Delta\lambda=77$\AA. To measure the continuum, a filter with 
$\lambda_{\rm cent}=6427$\AA, $\Delta\lambda=122$\AA \, was 
used.\footnote{See {https://arca.sai.msu.ru/filters?ics=NBI} for 
details.} For absolute calibration, a FITS image of the galaxy from 
\citet{james2004}, taken from the NED database, was used.

\section{METHODS OF RESEARCH}
\label{sect:method}

The research methods used in this study have been previously detailed in 
\citet{gusev2016,gusev2018,gusev2023}. The algorithm includes identifying 
SFRs using the SExtractor software in the $B$ and H$\alpha$ bands, 
determining the sizes of SFRs, their morphology, and their relationship 
with HII~regions, photometry of SFRs in individually selected apertures in 
$UBVRI$H$\alpha$, accounting for the contribution of gas emission in the 
$UBVRI$ bands, accounting for extinction in SFRs using spectral data 
(specifically the Balmer decrement), and estimating the mass and age of 
the stellar population in SFRs by comparing the luminosities and color 
indices of SFRs with a grid of evolutionary models calculated based on 
stellar evolutionary tracks library 
(version~2.8),\footnote{http://stev.oapd.inaf.it/cgi-bin/cmd/} 
developed by the Padova group \citep{bertelli1994,bressan2012,tang2014}. 
This grid corresponds to the metallicity on the color-luminosity and 
two-color diagrams, and the method involves finding the minimum of 
a deviation functional between observed values and the grid nodes. 
The deviation functional is defined as
\begin{eqnarray}
[[(U-B)_{\rm obs}-(U-B)_{\rm model}]^2 \\ \nonumber
+[(B-V)_{\rm obs}-(B-V)_{\rm model}]^2 \\
+[M(B)_{\rm obs}-M(B)_{\rm model}]^2]^{1/2}, \nonumber
\end{eqnarray}
where the subscript ''obs'' denotes observed luminosities and color 
indices, and ''model'' denotes model luminosities and color indices.

The result of this method is the determination of the following parameters 
of the SFRs: their sizes, gas metallicity, mass, and age of the stellar 
population. A detailed description of the resulting electronic catalog of 
SFRs is presented in \citet{gusev2023}. In this section, we focus on two 
key aspects of our methodology that posed the most significant challenges.

\subsection{Accounting for Extinction in SFRs}

The first key aspect concerns the correct accounting of extinction in the 
star formation region. Extinction is determined by the Balmer decrement 
of the gas emission in HII~regions. Using this value to account for 
extinction in the young stellar system in the $UBVRI$ bands is only possible 
if the stellar extinction equals the extinction in the gas cloud.

The evolutionary morphological sequence of SFRs \citep{whitmore2011} 
describes several stages of the development of a star-gas-dust region 
\citep[see][Fig. 1]{whitmore2011}. For young clusters that are unresolved 
into individual stars in the optical range, four development stages can 
be identified: \\
(1) emission in H$\alpha$ is observed, but no emission is seen in 
short-wavelength photometric bands; \\
(2) emission from both gas (in the H$\alpha$ line), and stars (in broad 
photometric bands) is observed, with the photometric centers of gas and 
star emissions coinciding; \\
(3) similar to stage~2, but the center of H$\alpha$ emission is shifted 
relative to the center of the stellar component emission; \\
(4) a blue stellar condensation is observed in $UBVRI$, but no H$\alpha$ 
emission is present \citep[see][Fig.~1 for illustration]{gusev2023}. The 
first stage corresponds to a very young, highly dusty region with significant 
internal extinction. As extinction decreases, the SFR transitions to the 
second stage. The third stage occurs when the gas shell expands due to the 
explosions of the first supernovae. In the fourth stage, ionized hydrogen in 
the expanding shell recombines and cools. The emission centers of stars and 
gas are considered shifted if the angular distance between them exceeds 
$0.5\arcsec$.

This evolutionary morphological sequence pertains to low-mass SFRs, such 
as star clusters and associations. In star complexes, which are conglomerates 
of OB associations and clusters, the situation is more complex due to the 
potential presence of several SFRs of different ages. For them, we can only 
speak of a ''photometric'' age, where the latest star formation burst 
primarily contributes to its calculation. Our goal is to select regions where 
extinction measured by the Balmer decrement is from the same part of the 
SFR from which the main photometric flux originates. This goal is achieved 
for both large-scale SFRs, such as star complexes, and star associations 
(clusters).

\begin{figure*}
\vspace{10mm}
\centerline{\includegraphics[width=17.8cm]{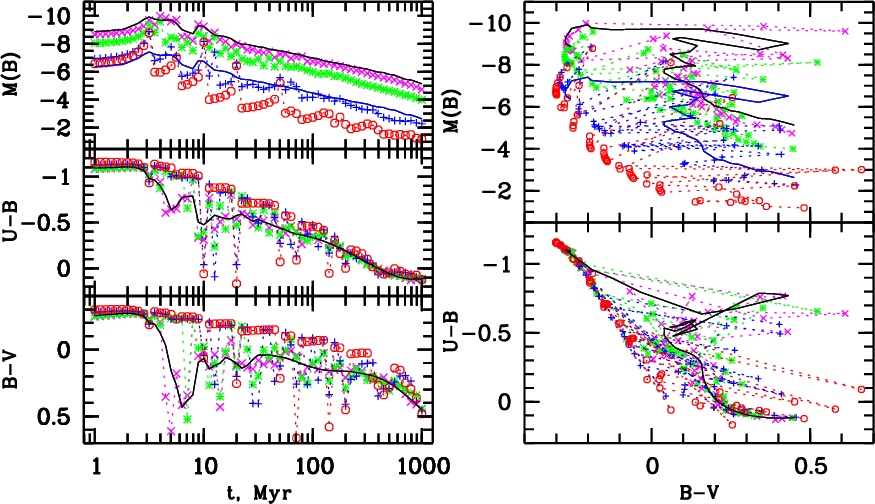}}
\caption{Examples of evolutionary sequences from the standard mode of SSP 
models (continuously populated IMF) from the Padova stellar evolutionary track 
library CMD version~2.8 \citep{bertelli1994,bressan2012,tang2014} for stellar 
systems with metallicity $Z=0.012$ and masses $1\cdot10^3 M\sun$ (blue curves) 
and $1\cdot10^4 M\sun$ (black curves) and the discrete mode of models (randomly 
populated IMF) for stellar systems with $Z=0.012$ and masses $500 M\sun$ (red 
circles connected by dashed lines), $1\cdot10^3 M\sun$ (blue crosses connected 
by dashed lines), $5\cdot10^3 M\sun$ (green stars connected by dashed lines) 
and $1\cdot10^4 M\sun$ (purple diagonal crosses connected by dashed lines). The 
age range varies from 1~Myr to 1~Gyr. Shown are changes in absolute magnitude 
$M(B)$ and color indices $U-B$ and $B-V$ as functions of age (left), as well 
as the color-magnitude and two-color diagrams (right) for the model evolutionary 
sequences. The color indices for the standard mode models are independent of 
mass (blue and black curves overlap on the corresponding plots).
\label{fig:model1}}
\end{figure*}

Our study focuses on SFRs in the last three stages, selecting only those
without H$\alpha$ emission (stage four) with a color index $(U-B)_0^i$, 
corrected for Galactic extinction and disk inclination extinction, not 
exceeding $-0.537^m$ \citep[see justification in][]{gusev2018}. However, 
the ''true'' extinction of a young stellar cluster and, consequently, its 
luminosity, colors, mass, and age can only be determined in the second and 
fourth stages. In the second stage (class~2 in the catalog), as shown in 
\citet{gusev2016,gusev2023}, extinction in the stellar system corresponds 
to extinction determined by the Balmer decrement. In the fourth evolutionary 
stage (class~0), excess extinction in the SFR is insignificant due to the 
dispersion of the gas-dust cloud; extinction in the star cluster (complex), 
which does not emit in H$\alpha$, is assumed to be $A_{\rm G}+A_{\rm in}$. In 
the third stage (class~1), the Balmer decrement provides extinction in the 
densest and brightest parts of the gas shell surrounding the young star 
cluster (complex). It was shown in \citet{gusev2016} that using the Balmer 
decrement to estimate stellar extinction in such objects gives an incorrect 
excess value $A$. Therefore, as shown below, we were able to estimate the 
mass for only 11 (age for 10) stellar clusters out of 33 SFRs with measured 
Balmer decrement \citep[one of the HII~regions in NGC~7292, studied 
in][covers two star clusters]{gusev2021}. Additional five SFRs, for which mass 
and age estimates of the stellar component were obtained, are in the fourth 
evolutionary stage (not emitting in H$\alpha$). Since chemical composition 
data for them are absent, we used the average metallicity at the corresponding 
galactocentric distance obtained in \citet{gusev2023b} when comparing them 
with evolutionary model grids.

\subsection{Accounting for Stochastic Effects in the Initial Mass
Function (IMF)}

\begin{figure*}
\vspace{10mm}
\centerline{\includegraphics[width=17.8cm]{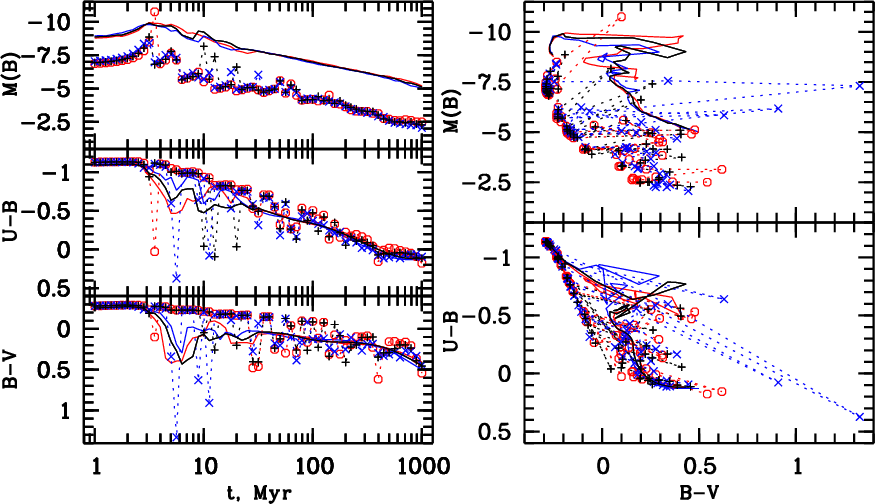}}
\caption{The same as in Fig.~\ref{fig:model1}, but for evolutionary sequences 
of varying metallicities. Examples include sequences from the standard mode 
models for stellar systems with mass $1\cdot10^4 M\sun$ and $Z=0.008$ (red 
curves), $Z=0.012$ (black curves), and $Z=0.018$ (blue curves), as well as 
from the discrete mode models for systems with mass $1\cdot10^3 M\sun$ and 
$Z=0.008$ (red circles connected by dashed lines), $Z=0.012$ (black crosses 
connected by dashed lines), and $Z=0.018$ (blue diagonal crosses connected 
by dashed lines).
\label{fig:model2}}
\end{figure*}

\begin{figure*}
\vspace{9mm}
\centerline{\includegraphics[width=17.8cm]{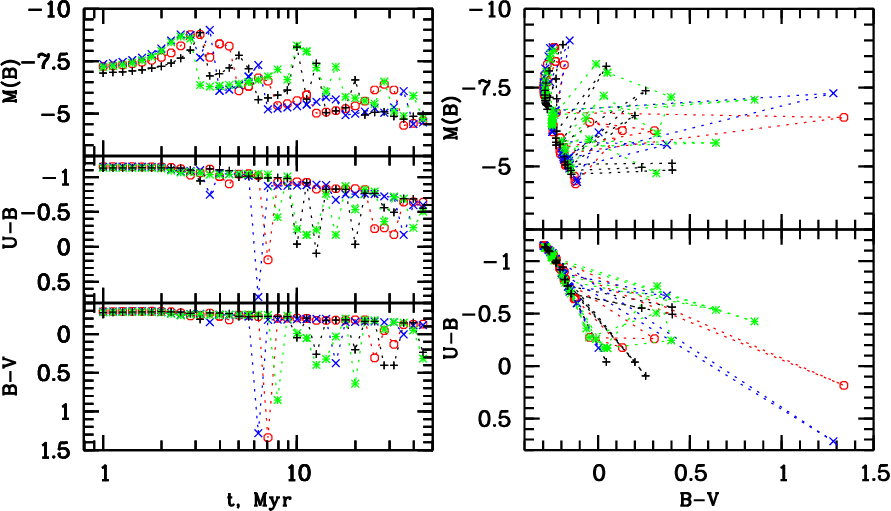}}
\caption{The same as in Fig.~\ref{fig:model1}, but for four randomly generated 
evolutionary sequences from the discrete mode of SSP models for systems with 
mass $1\cdot10^3 M\sun$ and $Z=0.012$ over an age range of 1 to 45~Myr.
\label{fig:model3}}
\end{figure*}

The second key aspect of the research methodology involves accounting for the 
stochastic effects in the discrete IMF, which is crucial for estimating the 
physical parameters of star clusters in SFRs. As demonstrated in 
\citet{whitmore2010,piskunov2011,cervino2013}, stochastic effects in a 
discretely and randomly populated IMF become significant for stellar systems 
with masses less than $5\cdot10^3-10^4 M\sun$. In these cases, the luminosity 
of short-lived red giants can become comparable to the total luminosity of 
main-sequence stars. The main issue here is the uncertainty in the number and 
masses of massive main-sequence stars (see 
Figs.~\ref{fig:model1}-\ref{fig:model3} for systems with a discrete IMF). 
Among the SFRs studied in this work, the effect of IMF discreteness is 
particularly important for young star clusters in the nearby galaxy NGC~7292.

A detailed description of the methodology for determining masses and ages 
using both continuously and stochastically populated IMFs is presented in 
\citet{gusev2023}. In this section, we will discuss unpublished results 
comparing the evolutionary sequences of stellar systems for randomly and 
continuously populated IMFs, and provide a brief description of our approach.

Modern evolutionary models of stellar populations allow us to determine the 
physical properties of stellar populations in individual young star clusters 
using detailed grids of spectral energy distributions (SEDs) and model colors, 
each corresponding to a unique age, metallicity, and IMF of the stellar 
population.

To avoid ambiguity in the age-metallicity relation, a separate grid of 
single stellar population (SSP) models was constructed for each object 
studied here, with a fixed metallicity determined from independent 
spectroscopic observations \citep{gusev2016}. Additionally, spectroscopic 
observations were used to calculate the Balmer decrement and estimate the 
interstellar extinction of the light emitted by the star cluster, allowing 
us to simultaneously account for the age--extinction relation that also 
affects the object's position on color-color diagrams and, consequently, 
the age estimation of the stellar population \citep{gusev2016}. The 
spectroscopic data enabled us to account for the influence of the emission 
from the surrounding interstellar gas on the broadband photometry of young 
stellar populations \citep{gusev2016}.

Taking these three factors into account allows for the comparison of the 
true integrated colors and lumi-nosities of star groupings with model 
predictions. This comparison is made with fixed values for metallicity, 
stellar light extinction, and the contribution of gas emission to the 
integrated spectra of star groupings, which are determined from independent 
spectral observations, rather than leaving these values as free parameters. 
This approach requires a combination of photometric, H$\alpha$ 
spectrophotometric, and spectroscopic observations of young star groupings.

The ages and masses of the star cluster populations were determined by 
minimizing the deviation function calculated for each node of the model grid, 
comparing the true (corrected for extinction and gas emission) colors and 
luminosities of the star cluster with the grid of model colors calculated for 
the fixed metallicity from observations.

The colors of bright, massive star clusters ($M>10^4 M\sun$) were compared 
with models computed using a continuously populated IMF. In the case of less 
massive star clusters ($M<10^4 M\sun$) stochastic effects due to the discrete 
IMF significantly affect the cluster's luminosity and color, especially in the 
long-wavelength part of the spectrum \citep{cervino2013}. Therefore, the 
colors and luminosities of low-mass star clusters were compared with models 
computed using a stochastically populated discrete IMF \citep{gusev2023}.

The discreteness of the IMF significantly affects the luminosity and colors 
of a cluster, which manifests as bursts and fluctuations in the evolutionary 
path of the photometric parameters of the cluster, caused by the appearance 
of red giants. There is also a systematic deviation between the photometric 
parameters (luminosity and color) of simple stellar population (SSP) models 
with discrete and continuous IMFs (Fig.~\ref{fig:model1}). The luminosity 
evolution curves of discrete models exhibit sloping oscillations and consist 
of relatively short time intervals of recurring events. During one time 
interval, there is a slow, gradual increase in the cluster's luminosity and 
an almost instantaneous burst, triggered by the evolution of the brightest 
main sequence star and its possible transformation into a bright, short-lived 
red supergiant. After the supergiant's explosion, the process repeats with 
the evolution of the next brightest main sequence star and its transformation 
into a red giant. It should be noted that the deviations in the color and 
luminosity evolution curves of the discrete model, described above, are 
stronger in the case of low-mass clusters, where the number of red giants is 
small, and the clusters spend most of the time as systems with main sequence 
stars. During bursts in low-mass clusters, the change in absolute stellar 
magnitude $B$ can exceed $2^m$, and the change in color can exceed $1^m$. 
Moreover, in the framework of the discrete model, young low-mass stellar 
clusters ($M\leq10^3 M\sun$ younger than 30~Myr) are systematically bluer by 
$\sim0.3-0.5^m$ in short-wavelength color indices than stellar systems of 
similar mass within the classical continuously populated IMF (see left graphs 
in Fig.~\ref{fig:model1}); on the color-luminosity diagram, they are located 
to the left of the evolutionary tracks of stellar systems with a continuously 
populated IMF, and on the two-color diagram, they are to the left and above 
(see right graphs in Fig.~\ref{fig:model1}). As the cluster mass increases 
to $1\cdot10^4 M\sun$, the color and luminosity evolution curves of the 
discrete model converge to the curves of the standard continuous model 
(Fig.~\ref{fig:model1}).

It should be noted that the differences between the model luminosities and 
color indices of stellar systems with different metallicities in the range 
of typical (solar and subsolar) $Z$ are insignificant (Fig.~\ref{fig:model2}) 
and generally do not exceed the characteristic measurement errors of the 
luminosities and color indices of real SFRs. Exceptions are the moments of 
bursts in low-mass stellar clusters (discrete IMF) and complexes with ages 
of 5-10~Myr (standard IMF). Extremely strong changes in the color indices 
$U-B$ and $B-V$ for a high-metallicity ($Z=0.018$) stellar system in the 
discrete IMF model in Fig.~\ref{fig:model2} (blue color) are not due to the 
large value of $Z$. A set of randomly generated models with a discrete IMF for 
systems with $Z=0.012$, presented in Fig.~\ref{fig:model3}, shows similar 
significant deviations for less metallic stellar systems. They occur in 
clusters and associations aged 5--10~Myr, at the same ages when the reddening 
and complex non-linear nature of the color index changes are also shown by 
models with a continuously populated IMF (Figs.~\ref{fig:model1}, 
\ref{fig:model2}).

To construct the SSP model grid with a discrete IMF, the Monte Carlo method 
was used, generating random changes in the discrete IMF depending on the 
given mass of the model stellar cluster. For each given cluster mass, 
a discrete IMF was generated using a random number generator. With this 
random sampling, for a fixed value of the stellar system's mass, the number 
of stars in the system was also fixed. At the next stage, using the randomly 
selected discrete IMF, the evolutionary sequence of 68 SSP models was 
calculated for each given cluster mass with a step of 0.05~dex in the 
interval $\log t=5.9-9.3$ and a fixed metallicity $Z$, obtained from 
observations. For each calculation of the randomly selected discrete IMF, 
a random initial number was used, also obtained through a random number 
generator. When comparing the observed colors and luminosity of a given object 
with the model, 50 evolutionary sequences of discrete SSP models were used in 
each iteration. The number of iterations per object varied from 2 to 4. Thus, 
the number of simulations of the discrete IMF with random sampling per object 
ranged from 50 to 200. The total number of modeled models for each pair of 
mass and age estimates of the object ranged from 6800 to 13600.

To calculate the errors in age and mass estimates for the case of discrete 
IMF models based on the model colors corresponding to the chosen grid node 
with the minimum deviation functional value, the corresponding interval on 
the color evolution curve of the discrete model between two ''red bursts'' 
was selected, and the interpolation polynomial coefficients were calculated. 
Then, knowing the functional relationship between age and color, as well as 
the observational errors of the color indices, we obtained the accuracy of 
age estimates using the Gaussian formula for calculating the error of indirect 
measurements. Similarly, using the functional correlation between the model 
luminosity and the cluster mass in the interval between two ''red bursts'' 
and the observational errors of the objects' integral luminosities, the mass 
estimate errors were determined.

A comparison of the mass and age estimates of GCs obtained using continuous 
and randomly populated IMFs for the same objects, conducted in \citet{gusev2023}, 
showed that the mass and age estimates obtained within the discrete model are 
systematically larger, which is consistent with the systematic excess of 
luminosities of continuous IMF models over the luminosities of discrete models 
with the same stellar population masses \citep[see][Figs.~5 and 6]{piskunov2011}. 
This difference decreases for high-mass ($M>5\cdot10^4 M\sun$) and very young 
($t<3$~Myr) or aging ($t>50$~Myr) stellar groups.

\begin{figure}
\vspace{5mm}
\centerline{\includegraphics[width=8cm]{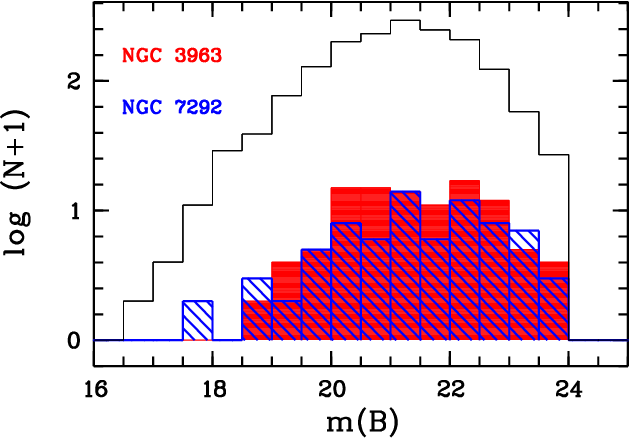}}
\caption{Luminosity function for the complete sample of SFRs in the catalog 
(black histogram), regions in NGC~3963 (red), and regions in NGC~7292 (blue).
\label{fig:bvis}}
\end{figure}

\section{RESULTS}

\subsection{Number of Identified SFRs, Their Sizes, and Locations in Galaxies}

We have identified a total of 157 SFRs: 93 in the galaxy NGC~3963 and 
64 in NGC~7292. Photometry was performed for all objects in the $UBVRI$ 
bands and in the H$\alpha$ (H$\alpha$ and [NII]$\lambda$6584 separately 
for SFRs in NGC~3963, and H$\alpha$+[NII] in NGC~7292), sizes were determined, 
and structures were described. Figure~\ref{fig:bvis} shows the distribution of 
the identified SFRs by apparent stellar magnitude in the $B$ band in comparison 
with the full sample of objects in the catalog. The number of studied SFRs in 
each galaxy is insufficient to determine the slope of the luminosity function 
and assess the completeness of the object sample. Assuming that the luminosity 
function follows a power law, we can qualitatively assess the completeness of 
the sample by the distribution peak of the regions in Fig.~\ref{fig:bvis}. 
Roughly, we estimate that our samples for NGC~3963 and NGC~7292 are complete 
at least up to $m(B)=20^m$, which approximately corresponds to the limiting 
magnitudes to which the SFR samples are complete for most catalog galaxies 
\citep[Fig.~\ref{fig:bvis}, see also][Fig.~4 therein]{gusev2023}.

As explained in Section~\ref{sect:method}, estimating the mass and age of the 
stellar populations in the regions using our method is possible only for 
class~0 regions and those class~2 regions for which spectroscopic data are 
available. The number of SFRs satisfying these criteria is 16, with eight in 
each galaxy.

\begin{figure*}
\vspace{9mm}
\centerline{\includegraphics[width=17.8cm]{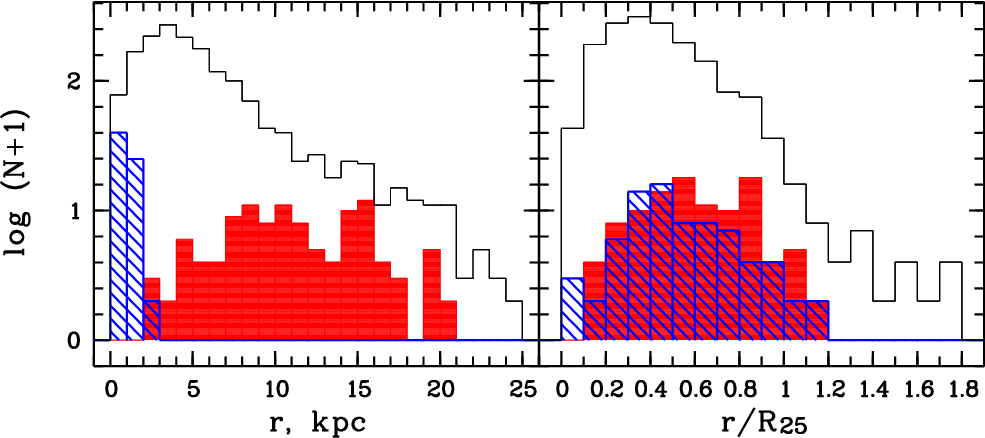}}
\caption{Distribution of SFRs by absolute (left) and normalized to $R_{25}$ 
(right) galactocentric distances. Notations are the same as in Fig.~\ref{fig:bvis}.
\label{fig:rad}}
\end{figure*}

The galaxy NGC~7292 is the smallest in linear size among all 21 galaxies in 
the catalog. Consequently, all regions in it are located at galactocentric 
distances $r<2$~kpc, comprising a significant portion of the regions close to 
the center of the entire catalog (see the left graph in Fig.~\ref{fig:rad}). 
In the large, massive NGC~3963, the SFRs are distributed over a wide range $r$, 
mainly concentrating in the spiral arms of the galaxy (Figs.~\ref{fig:map}, 
\ref{fig:rad}).

Despite significant differences in absolute values of $r$, the distributions 
of the regions by galactocentric distance normalized to the optical radius 
$R_{25}$ in both galaxies are almost identical (right graph in 
Fig.~\ref{fig:rad}), completely covering the range $r$ within $R_{25}$.

\begin{figure}
\vspace{4mm}
\centerline{\includegraphics[width=8cm]{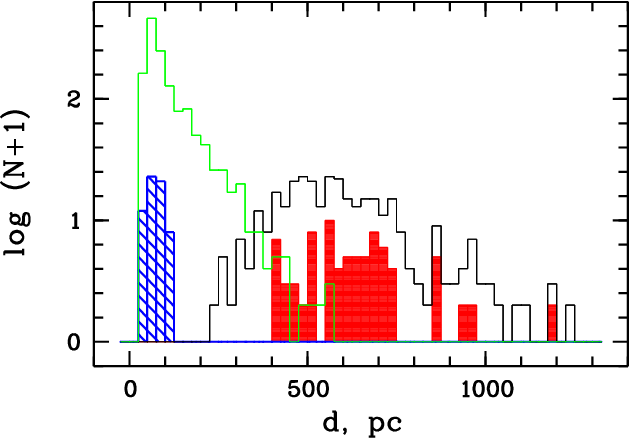}}
\caption{Distribution of SFRs by size for objects in galaxies from the 
catalog located within 30~Mpc (green histogram) and beyond 30~Mpc from us 
(black histogram). Other notations are the same as in Fig.~\ref{fig:bvis}.
\label{fig:size}}
\end{figure}

Since NGC~3963 and NGC~7292 are located at substantially different distances 
from us (see Table~\ref{tab:gen}), due to the hierarchical processes of star 
formation, we observe young regions in the nearby NGC~7292 with diameters of 
40--125~pc (Fig.~\ref{fig:size}). The average value of $d=75\pm20$~pc for the 
SFRs in NGC~7292 corresponds to the characteristic sizes of stellar associations 
\citep{efremov1987,ivanov1991}. The sizes of the SFRs in NGC~3963, ranging from 
400 to 750~pc with an average of $d=650\pm200$~pc (Fig.~\ref{fig:size}), are 
typical for stellar complexes \citep{efremov1987,ivanov1991}. It is worth noting 
that the largest SFRs in NGC~3963 with $d>800$~pc are diffuse objects of 
relatively low surface brightness, possibly consisting of several complexes.

\subsection{Gas Metallicity, Contribution of Its Emission to Total Luminosity in 
            Broad Photometric Bands, and Morphology of Gas and Stellar Emission 
            in SFRs}

\begin{figure}
\vspace{4mm}
\centerline{\includegraphics[width=8cm]{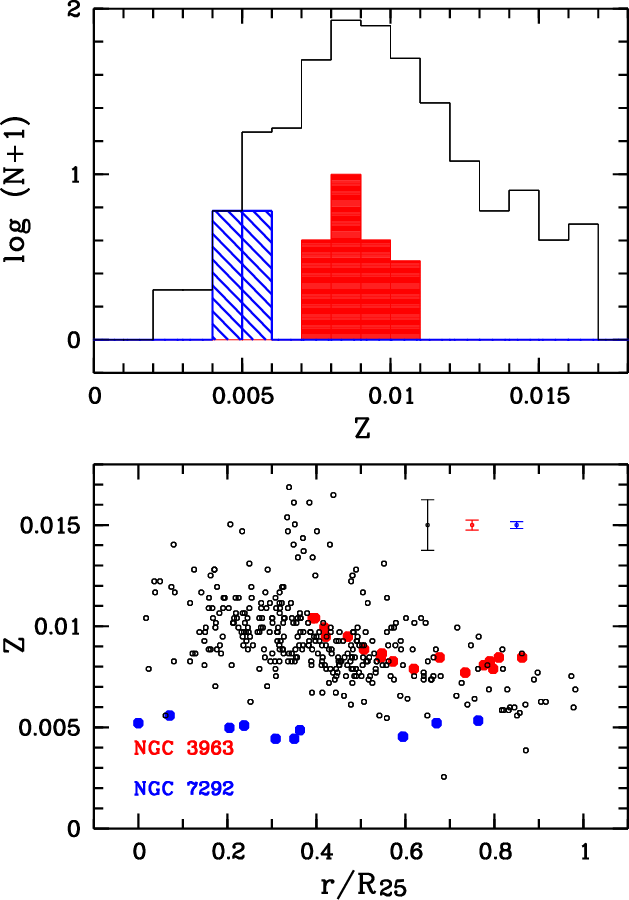}}
\caption{Distribution of SFRs by metallicity (top; notations are the 
same as in Fig.~\ref{fig:bvis}) and radial distribution of SFRs by $Z$ 
(bottom). Small black circles on the lower graph represent the complete sample 
of SFRs except for those with metallicity determined using the HII-ChiMistry 
method, red circles indicate SFRs in NGC~3963, and blue circles indicate SFRs 
in NGC~7292. Average measurement errors for $Z$ in the SFR samples are shown 
in corresponding colors.
\label{fig:z}}
\end{figure}

Differences in the masses of NGC~3963 and NGC~7292 significantly affect the 
variations in the chemical composition of their SFRs. The SFR metallicities were 
determined in our study \citet{gusev2023} using several ''strong line'' 
methods (R \citep{pilyugin2016}, S \citep{pilyugin2016}, O3N2 \citep{pettini2004}, 
NS \citep{pilyugin2011}, and HII-ChiMistry \citep{perez2014}). For the HII~regions 
in NGC~3963 and NGC~7292, the R~method (using the lines [OII], [OIII], and [NII]) 
and the S~method (using the lines [OIII], [NII], and [SII]) were employed 
\citep{gusev2021}. The SFR metallicity, $Z$, was determined from the oxygen 
abundance, O/H, as the arithmetic mean of the values obtained by both methods. 
In the giant massive galaxy NGC~3963, the SFR metallicity is average for the 
entire catalog sample, with its central metallicity being one of the highest 
among the 21 galaxies in the sample (Fig.~\ref{fig:z}). In contrast, the small, 
low-mass NGC~7292 has the lowest metallicity of SFRs among all catalog galaxies 
(Fig.~\ref{fig:z}). It is also typical for irregular galaxies to lack a 
metallicity gradient \cite[e.g.,][]{richer1995,hernandez2009}.

It should be noted that the $Z$ values obtained using all methods are consistent 
within the measurement errors \citep{gusev2023}, except for the HII-ChiMistry 
method, which systematically overestimates $Z$ 
\cite[see][Fig.~7 therein]{gusev2023}. Therefore, the chemical composition 
parameters of the gas in the SFRs of NGC~3963 and NGC~7292 are not influenced 
by the calibration specifics of the methods used.

\begin{figure}
\vspace{4mm}
\centerline{\includegraphics[width=8cm]{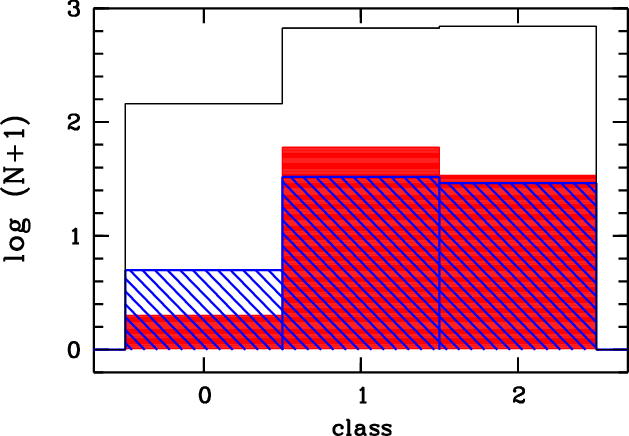}}
\caption{Distribution of SFRs by gas and stellar emission morphology (see 
Section~\ref{sect:method}) for the complete sample of SFRs in the catalog 
galaxies with available H$\alpha$ line observations (classes~0-2; black 
histogram), SFRs in NGC~3963 (red), and SFRs in NGC~7292 (blue).
\label{fig:hadata}}
\end{figure}

In Section~\ref{sect:method}, we highlighted the importance of studying the 
morphology of gas and stellar component emission in the H$\alpha$ line and 
broad photometric bands. Figure~\ref{fig:hadata} shows the distribution 
of the studied SFRs by morphological classes (evolutionary 
stages).\footnote{In \citet{gusev2023}, 341 class~2 objects with no Balmer 
decrement measurements were erroneously assigned to class~1. In this paper, 
we correct this error.} As seen, the number of regions with coinciding and 
offset centers of gas emission and stellar radiation is approximately equal 
in both NGC~3963 and NGC~7292, as well as in the full sample. Given that our 
method for detecting young stellar groups and HII~regions is independent 
of the morphological class of the SFRs (classification into classes occurs 
in subsequent stages), the approximate equality in the number of class~1 
and class~2 regions is not a selection effect but reflects the roughly 
equal time spent by young stellar regions in these evolutionary stages. The 
number of class 0~regions (without H$\alpha$ emission) is $11\%$ for the 
full sample of regions with known H$\alpha$ data (class~0–2 regions). In 
the galaxies studied here, there are only four such regions ($6\%$) in 
NGC~7292 and one ($1\%$) in NGC~3963 (Fig.~\ref{fig:hadata}).

\begin{figure}
\vspace{9mm}
\centerline{\includegraphics[width=6cm]{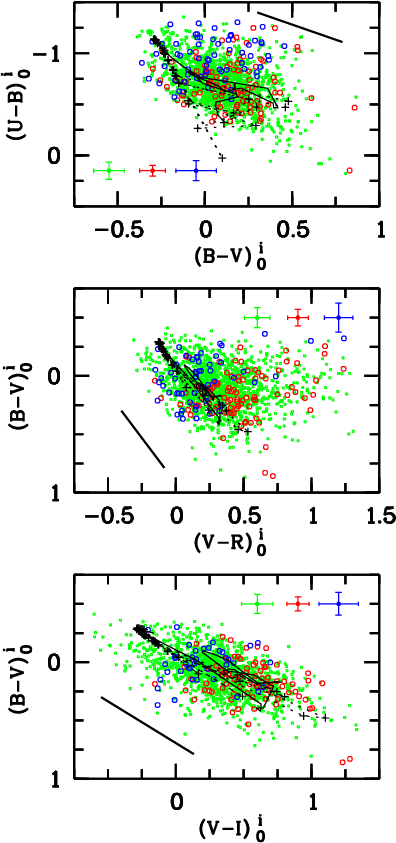}}
\caption{Two-color diagrams $(U-B)_0^i-(B-V)_0^i$, $(B-V)_0^i-(V-R)_0^i$, 
and $(B-V)_0^i-(V-I)_0^i$. Shown are SFRs with color indices corrected 
for galactic extinction $A_{\rm G}$ and extinction caused by the galaxy 
disk inclination, $A_{\rm in}$. Green dots represent SFRs from the 
catalog by \citet{gusev2023}, red circles denote SFRs in NGC~3963, and 
blue circles denote SFRs in NGC~7292. Average measurement errors for 
color indices are shown with corresponding color error bars. Black curves 
represent evolutionary tracks for a stellar system with a continuous IMF 
and $Z=0.008$ in the age range $t$ of 1 to 100~Myr; black crosses 
connected by dashed lines show an example of an evolutionary sequence 
for a stellar system with a randomly populated IMF with mass 
$M=5\cdot10^3 M\sun$ and $Z=0.008$ in the range $t$ from 1 to 100~Myr. 
Black thick segments in the corners of the diagrams are parallel to the 
reddening vector.
\label{fig:color}}
\end{figure}

\citet{gusev2023} found that a significant (more than $10\%$) portion of 
SFRs has a gas emission contribution exceeding $40\%$ in the photometric 
$B$ band. Among the regions in NGC~3963 and NGC~7292, no objects with 
such a high gas emission contribution were found: in most SFRs, this 
contribution does not exceed $15\%$. However, in the SFRs of NGC~7292, 
the gas emission contribution in the $B$ band (average $12\%$, maximum 
$35\%$) is systematically higher than in the SFRs of NGC~3963 (average 
$5\%$, maximum $22\%$). For class~2 regions, the data on gas contribution 
are presented in Table~\ref{tab:phot}.

\subsection{Photometric Parameters of SFRs}

\begin{figure}
\vspace{9mm}
\centerline{\includegraphics[width=6cm]{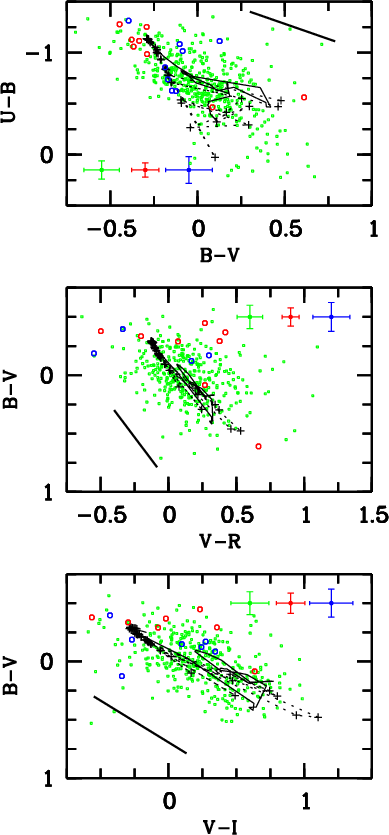}}
\caption{The same as in Fig.~\ref{fig:color}, but for the ''true'' color 
indices of the stellar population in SFRs corrected for the contribution of 
gas and extinction calculated from the Balmer decrement. Notations are the same 
as in Fig.~\ref{fig:color}.
\label{fig:color2}}
\end{figure}

The positions of the studied SFRs on two-color diagrams are shown in 
Figs.~\ref{fig:color} and \ref{fig:color2}. Figure~\ref{fig:color} 
presents the color indices of the SFRs corrected for Galactic extinction 
and extinction associated with the inclination of the galactic disks. 
Although the actual extinction in SFRs typically exceeds the sum of 
extinctions $A_{\rm G}+A_{\rm in}$ \citep{gusev2023}, and accounting 
for gas contribution can shift SFRs away from the reddening vector 
\citep[see][Fig. 12 therein]{gusev2016}, the majority of catalog objects 
are compactly located on the diagrams -- within areas corresponding to 
young stellar systems with continuous or discrete star formation 
histories (Fig.~\ref{fig:color}). An exception is the $(B-V)_0^i-(V-R)_0^i$ 
diagram, where SFRs with very high color indices $(V-R)_0^i$ up to $1.2-1.3$ 
are observed. This is due to the significant contribution of gas to the 
flux in the $R$ band \citep{gusev2023}. The objects in NGC~3963 and NGC~7292 
studied in this work do not stand out among the SFRs of the full catalog 
sample on the diagrams in Fig.~\ref{fig:color}.

\begin{figure}
\vspace{5mm}
\centerline{\includegraphics[width=8cm]{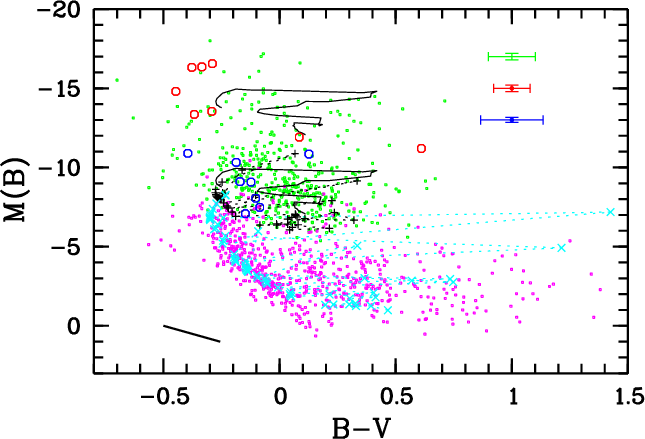}}
\caption{The color-luminosity diagram for the ''true'' absolute stellar 
magnitudes $M(B)$ and color indices $B-V$ of the stellar population in SFRs 
in the galaxies. Purple points represent open star clusters in our Galaxy from 
the catalog of \citet{kharchenko2009}. Cyan crosses connected by dashed lines 
show an example of an evolutionary sequence for a stellar system with a randomly 
populated IMF with mass $M=500 M\sun$ and $Z=0.018$ in the range $t$ from 1~Myr 
to 1~Gyr. Black curves indicate the evolutionary tracks for a stellar system 
with a continuous IMF, with masses $M=1\cdot10^6 M\sun$ (upper curve) and 
$M=1\cdot10^4 M\sun$ (lower curve) and $Z=0.008$ in the age range $t$ from 
1 to 100~Myr. Other notations are the same as in Fig.~\ref{fig:color}.
\label{fig:cmd}}
\end{figure}

On the diagrams in Fig.~\ref{fig:color2}, where the colors of objects are 
corrected for gas contribution and extinction obtained from the Balmer 
decrement, the cloud of points representing color indices also concentrates 
along model evolutionary tracks. The degree of concentration is lower than 
on the diagrams in Fig.~\ref{fig:color}, which is explained by the smaller 
number of objects (only class~0 and class~2 SFRs with known Balmer decrement) 
and larger measurement errors of the ''true'' color indices. On the 
$(B-V)-(V-R)$ diagram, the group of objects with extremely high $V-R$ values 
disappears, which may indicate correct accounting of gas contribution to 
the fluxes in the photometric bands.

It is noteworthy that on the $(U-B)-(B-V)$ diagram, most SFRs from NGC~3963 
are located in the upper left corner of the diagram representing the region 
of the youngest stellar populations. In contrast, half of the eight regions 
in the nearby galaxy NGC~7292 have color indices corresponding to stellar 
systems with discretely populated star formation histories during periods 
without outbursts (stars in the stage of red (super)giants; 
Figs.~\ref{fig:model1}-\ref{fig:model3}, \ref{fig:color2}).

As indicated in Section~\ref{sect:method}, the physical parameters of SFRs 
were evaluated using two color indices: $U-B$ and $B-V$. SFRs in the $R$ and 
$I$ bands are distinguished much worse than in the shorter wavelength bands. 
In some cases, we were even unable to obtain photometry with satisfactory 
accuracy, which is why the number of SFRs on the $(B-V)-(V-R)$ and 
$(B-V)-(V-I)$ diagrams is lower than on the $(U-B)-(B-V)$ diagram 
(Fig.~\ref{fig:color2}). In addition, fluxes in $R$ and $I$ are weakly 
sensitive to changes in age in young stellar systems and, conversely, are 
more sensitive to the possible presence of old stellar populations (e.g., 
due to non-simultaneous or repeated star formation outbursts, poorly described 
by SSP models). For fluxes in the $R$ band, another factor is the large 
contribution of gas emission to the total flux. Errors in spectroscopic data 
give additional uncertainty to measurements in $R$. Consequently, there is 
a significant lack of agreement with models for a considerable number of 
studied SFRs on the $(B-V)-(V-R)$ and $(B-V)-(V-I)$ diagrams 
(Fig.~\ref{fig:color2}). It is also noted that stochastic effects in star 
formation histories manifest in longer wavelength bands for larger stellar 
systems compared to shorter wavelength bands $U$, $B$, $V$ \citep{cervino2013}.

\begin{figure}
\vspace{5mm}
\centerline{\includegraphics[width=8cm]{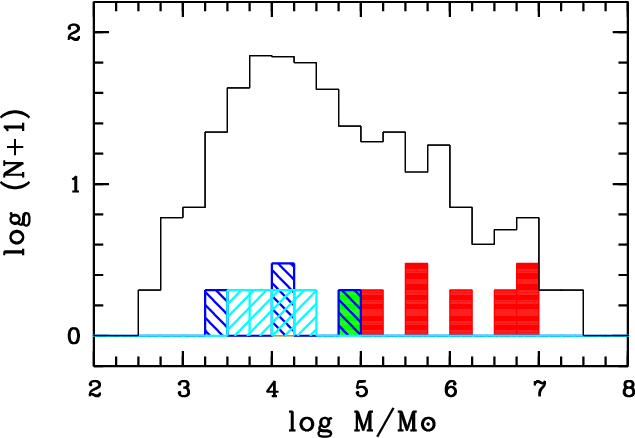}}
\caption{Distribution of SFRs by mass for the complete catalog sample (black 
histogram), class~2 SFRs in NGC~3963 (red), class~0 SFRs in NGC~3963 (green), 
class~2 SFRs in NGC~7292 (blue), and class~0 SFRs in NGC~7292 (cyan).
\label{fig:mass}}
\end{figure}

\begin{figure}
\vspace{5mm}
\centerline{\includegraphics[width=8cm]{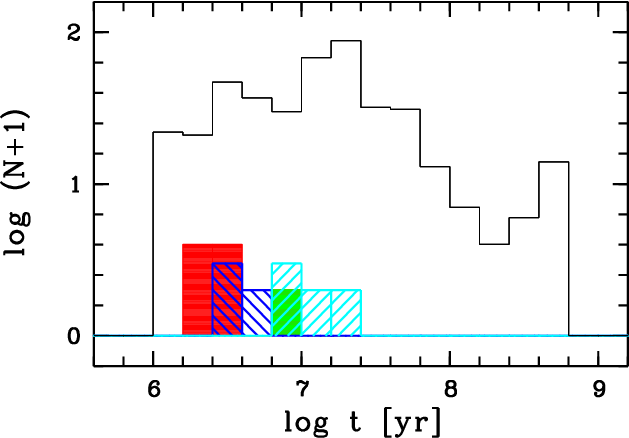}}
\caption{Distribution of SFRs by age. Notations are the same as in 
Fig.~\ref{fig:mass}.
\label{fig:tage}}
\end{figure}

On the color-luminosity diagram (Fig.~\ref{fig:cmd}), in addition to the objects 
in our catalog, we show the positions of open star clusters in our Galaxy from 
catalog of \citet{kharchenko2009}. SFRs in NGC~3963 and NGC~7292 do not stand 
out from the objects in the full catalog sample on this diagram. We observe 
high-mass stellar complexes in the distant NGC~3963 as expected, while in the 
nearby NGC~7292, relatively low-mass stellar associations are observed 
(Fig.~\ref{fig:cmd}). The least massive SFRs in NGC~7292 correspond in luminosity 
to the most massive open star clusters in the Galaxy (Fig.~\ref{fig:cmd}).

\begin{figure*}
\vspace{9mm}
\centerline{\includegraphics[width=17.8cm]{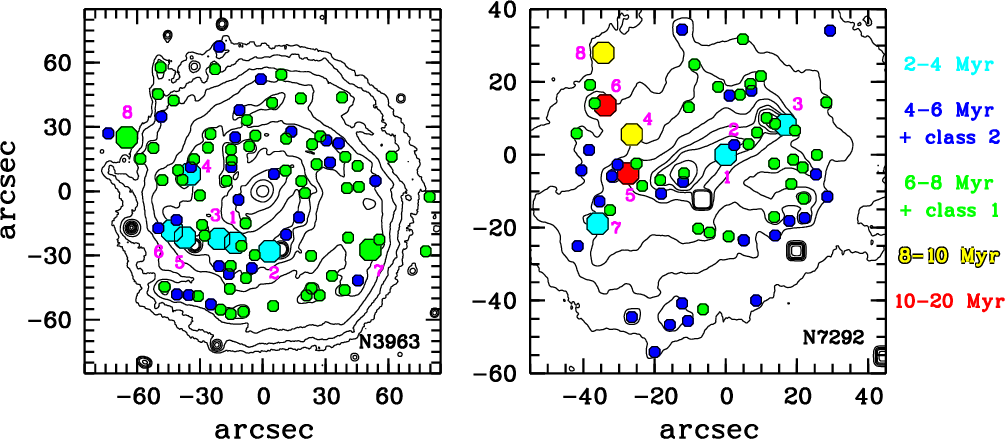}}
\caption{Distribution of SFRs of different ages and evolutionary classes in the 
galaxies NGC~3963 (left) and NGC~7292 (right). Large circles denote SFRs with 
measured age estimates, while small circles represent other SFRs of classes~1 
and~2. Purple numbers correspond to the SFRs listed in Table~\ref{tab:phys}.
\label{fig:tmap}}
\end{figure*}

\subsection{Estimates of the Mass and Age of the Stellar Component of 
            Star Formation Regions}
            
The range of stellar population masses in SFRs in the galaxy NGC~7292, 
$M=3\cdot10^3 - 6\cdot10^4 M\sun$, corresponds to typical masses for objects in 
the catalog (Fig.~\ref{fig:mass}). The most massive complexes in NGC~3963 reach 
$8\cdot10^6 M\sun$, making them some of the most massive SFRs among all objects 
in the catalog (Fig.~\ref{fig:mass}). Only the largest complexes in NGC~2336, 
NGC~5351, and NGC~7678 have larger masses \citep{gusev2023}.

It should be noted that SFRs without emission in the H$\alpha$ line (class~0) are 
generally less massive in both galaxies (in NGC~7292, class~0 objects have an 
average mass $M=(1.0\pm0.7)\cdot10^4 M\sun$ with a median value 
$M=0.64\cdot10^4 M\sun$, compared to an average mass 
$(2.3^{+2.4}_{-2.3})\cdot10^4 M\sun$ and median mass of $1.4\cdot10^4 M\sun$ for 
class~2 objects; the single class~0 object in NGC~3963 was the least massive 
among the eight SFRs with measured mass). This can be explained by selection 
effects: for comparable apparent magnitudes, the extinction in regions with gas 
emission is on average higher by $\sim2^m$ in the $B$ band. A small decrease in 
luminosity with age does not compensate for the effect of excessive extinction. 
An exception is the least massive HII~region in NGC 7292, which also has the 
highest $m(B)$ value among the eight regions with mass estimates in the galaxy. 
This object was not identified by the SExtractor software and was included in 
our sample due to a more thorough search in photometric bands for HII~regions 
studied in \citet{gusev2021}.

Among the 16 SFRs with mass estimates, the age of the stellar population could 
be determined for 15. For one region in NGC~7292, the error in age $t$ 
estimation was too large for any qualitative analysis. The distribution of SFRs 
by age is shown in Fig.~\ref{fig:tage}. With a full range of measured ages from 
2.2 to 18~Myr, only two regions in NGC~7292 have $t>10$~Myr. On average, 
complexes in NGC~3963 show younger ages: six out of eight SFRs in the galaxy 
are younger than 3~Myr, and the other two are younger than 8~Myr. In NGC~7292, 
we did not find any stellar associations younger than 3.1~Myr.

The difference in ages of SFRs is due to the fact that a significant proportion 
of studied objects in NGC~7292 do not have emission in H$\alpha$ (are not 
HII~regions). As mentioned in Section~\ref{sect:method}, classes~0 and 2 
represent different evolutionary stages of SFRs. The distribution in 
Fig.~\ref{fig:tage} clearly shows this: the ages of HII~regions range from 
2.2 to 7.9~Myr, while regions without gas emission have ages $t$ from 6.3 to 
18~Myr. This separation provides additional support for the validity of our 
age $t$ estimates.

The distribution of SFRs of different ages in galaxies is shown in 
Fig.~\ref{fig:tmap}. The age of objects for which we could not estimate $t$ can 
be roughly inferred from their morphological class -- a marker of evolutionary 
development of SFRs (see Section~\ref{sect:method}). The age boundary between 
classes~1 and 2 is approximately 4–5~Myr \citep{whitmore2011}. Such objects are 
marked with small circles in Fig.~\ref{fig:tmap}, and their color approximately 
corresponds to the color of SFRs of the corresponding age (see the legend on 
the right side of the figure).

\begin{table*}
\begin{center}
\caption{Photometric parameters of SFRs}
\label{tab:phot}
\begin{tabular}{c|c|c|c|c|c|c|c}
\hline\hline
$N$ & NGC & Coordi- & $m(B)$, & $M(B)$, & $U-B$ & $B-V$ & Gas \\
    &     & nates   & mag     & mag  &    &    & contribution \\
\hline
1528 & 3963 & 13.1$\arcsec$E, 24.1$\arcsec$S & 
$21.40\pm0.02^m$ & $-14.90\pm0.09^m$ & $-1.28\pm0.03$ 
& $-0.45\pm0.04$ & 0.14 \\
1531 & 3963 &  3.3W, 28.0S & $20.05\pm0.03$ & $-16.56\pm0.06$ 
& $-0.98\pm0.04$ & $-0.29\pm0.04$ & 0.06 \\
1532 & 3963 & 20.9E, 22.3S & $21.30\pm0.02$ & $-16.32\pm0.23$ 
& $-1.13\pm0.05$ & $-0.38\pm0.06$ & 0.03 \\
1540 & 3963 & 34.8E, 8.0N  & $23.30\pm0.07$ & $-13.35\pm0.24$ 
& $-1.06\pm0.10$ & $-0.37\pm0.12$ & 0.22 \\
1556 & 3963 & 37.2E, 21.7S & $20.27\pm0.02$ & $-16.35\pm0.06$ 
& $-1.12\pm0.03$ & $-0.33\pm0.03$ & 0.08 \\
1564 & 3963 & 43.4E, 18.4S & $22.57\pm0.13$ & $-13.54\pm0.86$ 
& $-1.25\pm0.23$ & $-0.29\pm0.28$ & 0.12 \\
1576 & 3963 & 51.3W, 27.4S & $21.69\pm0.03$ & $-11.91\pm0.07$ 
& $-0.46\pm0.04$ &  $0.08\pm0.05$ & 0.03 \\
1597 & 3963 & 64.9E, 25.2N & $22.52\pm0.01$ & $-11.20\pm0.01$ 
& $-0.56\pm0.02$ &  $0.61\pm0.02$ & 0    \\
1604 & 7292 &  0.0, 0.0    & $20.53\pm0.14$ & $-10.32\pm0.20$ 
& $-0.86\pm0.19$ & $-0.19\pm0.21$ & 0.09 \\
1605 & 7292 &  2.4W, 2.7N  & $21.28\pm0.14$ & $-10.89\pm0.34$ 
& $-1.31\pm0.19$ & $-0.40\pm0.18$ & 0.06 \\
1617 & 7292 & 16.8W, 8.3N  & $19.78\pm0.04$ & $-10.85\pm0.05$ 
& $-1.11\pm0.06$ &  $0.13\pm0.05$ & 0.35 \\
1637 & 7292 & 26.4E, 5.6N  & $22.05\pm0.09$ &  $-7.48\pm0.09$ 
& $-1.02\pm0.10$ & $-0.09\pm0.12$ & 0    \\
1642 & 7292 & 27.5E, 5.1S  & $20.43\pm0.05$ &  $-9.10\pm0.05$ 
& $-0.74\pm0.06$ & $-0.17\pm0.08$ & 0    \\
1647 & 7292 & 33.9E, 13.6N & $22.45\pm0.11$ &  $-7.08\pm0.11$ 
& $-0.63\pm0.19$ & $-0.15\pm0.16$ & 0    \\
1657 & 7292 & 36.0E, 18.9S & $22.86\pm0.19$ &  $-8.06\pm0.25$ 
& $-1.08\pm0.23$ & $-0.10\pm0.23$ & 0.24 \\
1658 & 7292 & 34.4E, 28.0N & $20.46\pm0.01$ &  $-9.07\pm0.01$ 
& $-0.62\pm0.02$ & $-0.12\pm0.03$ & 0    \\
\hline
\end{tabular}
\end{center}
\end{table*}

\begin{table}
\begin{center}
\caption{Physical and chemical parameters of SFRs}
\label{tab:phys}
\begin{tabular}{c|c|c|c|c|c}
\hline\hline
NGC & No. & $Z$ & $d$, pc, & $M$, $10^4 M\sun$ & $t$, Myr \\
\hline
3963 & 1 & 0.010 & 520 & $180\pm10$    & $2.2\pm2.1$ \\
3963 & 2 & 0.010 & 700 & $540\pm10$    & $2.8\pm0.4$ \\
3963 & 3 & 0.009 & 640 & $730\pm10$    & $2.2^{+3.2}_{-2.2}$ \\
3963 & 4 & 0.009 & 420 & $37\pm2$      & $2.5^{+2.8}_{-2.5}$ \\
3963 & 5 & 0.008 & 850 & $800\pm40$    & $2.2^{+2.3}_{-2.2}$ \\
3963 & 6 & 0.008 & 570 & $34\pm2$      & $2.8\pm2.4$ \\
3963 & 7 & 0.008 & 660 & $14\pm1$      & $7.9\pm1.0$ \\
3963 & 8 & --    & 560 & $6.3\pm0.5$   & $6.3\pm0.2$ \\
7292 & 1 & 0.005 &  75 & $1.4\pm0.2$   & $3.2\pm0.7$ \\
7292 & 2 & 0.006 &  75 & $5.8\pm0.4$   & -- \\
7292 & 3 & 0.004 &  65 & $1.7\pm0.1$   & $3.2\pm0.1$ \\
7292 & 4 & --    &  80 & $0.64\pm0.04$ & $8.9\pm0.5$ \\
7292 & 5 & --    &  90 & $2.0\pm0.1$   & $13\pm3$ \\
7292 & 6 & --    &  90 & $0.35\pm0.04$ & $18\pm4$ \\
7292 & 7 & 0.005 & 115 & $0.26\pm0.04$ & $4.0\pm0.1$ \\
7292 & 8 & --    & 125 & $1.00\pm0.04$ & $8.9\pm0.2$ \\
\hline
\end{tabular}
\end{center}
\end{table}

In the galaxy NGC~3963, SFRs younger than 6~Myr are mostly located in the 
spiral arms of the inner disk. In the outer part of the distorted southern 
spiral (Figs.~\ref{fig:map}, \ref{fig:tmap}), which is apparently influenced 
by the neighboring galaxy NGC~3958 \citep{moorsel1983}, the majority of 
complexes belong to class~1, and the only complex with an estimated has an 
age of 7.9~Myr.

In NGC~7292, the youngest regions ($t<4$~Myr or class~2) are located in the 
center of the bar, near the ends of the bar (but not in the brightest regions 
at the ends), as well as on the southern and eastern outskirts of the disk 
(Fig.~\ref{fig:tmap}). It is noted that in large stellar complexes both at the 
eastern and western ends of the bar, relatively older bright SFRs are located 
in the center of the stellar complex (green circles on the map), while younger 
SFRs are on its periphery (cyan and blue circles). Within stellar complexes in 
the galaxy, reaching 350-400~pc in diameter, close SFRs with relatively large 
age differences, up to 10-12~Myr, are observed. This results from 
non-simultaneous star formation in large stellar complexes. The typical 
duration of star formation on scales of $\sim300$~pc is about 20~Myr 
\citep[see][Fig.~8 therein]{efremov1998}. In NGC~5585, which is similar in 
mass and luminosity to NGC~7292 and has low surface brightness, the typical age 
difference between SFRs located 100~pc apart is also $\approx10$~Myr 
\citep[see][Fig.~9 therein]{gusev2019}. The brightest dynamically prominent 
HII~region in NGC~7292, located at the eastern end of the bar and previously 
thought to be the center of the galaxy, was examined by us in detail 
\citep[see][Fig.~8 therein]{gusev2023b}. Its age was estimated in 
\citet{gusev2023b} to be 6-8~Myr.

Tables~\ref{tab:phot} and \ref{tab:phys} provide information on the 16 SFRs in 
the galaxies for which physical parameters of the stellar population were 
estimated. Table~\ref{tab:phot} lists the catalog number $N$, coordinates in 
arcseconds relative to the center, apparent ($m(B)$) and ''true'' absolute 
stellar magnitudes ($M(B)$), ''true'' color indices $U-B$ and $B-V$, and the 
contribution of gas to the total flux in the band. Table~\ref{tab:phys} 
presents the object number from Fig.~\ref{fig:tmap}, gas metallicity $Z$, 
SFR size $d$, mass $M$, and age $t$ of the stellar population. A gas 
contribution of 0 in Table~\ref{tab:phot} and absence of data on $Z$ in 
Table~\ref{tab:phys} indicate the lack of emission in the H$\alpha$ line 
(evolutionary class~0). Complete data on the parameters of these and other 
141 SFRs in NGC~3963 and NGC~7292 are provided in version~3 of our 
catalog.\footnote{http://lnfm1.sai.msu.ru/$\sim$gusev/sfr\_cat.html}

\section{DISCUSSION OF THE RESULTS}

In this section, we will examine some general dependences for all objects in the 
catalog, which have both fundamental and methodological significance.

One possible indicator of the age of HII~regions is the equivalent width of the 
Balmer hydrogen emission lines \citep{copetti1986}. In \citet{gusev2018}, we 
introduced the index $R-{\rm H}\alpha = R+2.5\log F({\rm H}\alpha+[{\rm NII}])$, 
where $R$ is the apparent stellar magnitude in the $R$ band, and 
$F({\rm H}\alpha+[{\rm NII}])$ is the flux in erg$\cdot$c$^{-1}\cdot$cm$^{-2}$, 
which can serve as a replacement for EW(H$\alpha$) 
\citep[see][Fig.~17 therein]{gusev2023}. It should be noted that for NGC~3963 -- 
the only galaxy out of the 21 in the catalog for which separate flux measurements 
were made for the H$\alpha$ and [NII]$\lambda$6584 lines, we calculated the 
$R-{\rm H}\alpha$ index as 
$R+2.5\log [F({\rm H}\alpha)+F(1.33[{\rm NII}]\lambda6584)]$, where coefficient 
1.33 reflects the contribution of emission in the [NII]$\lambda$6548 line.

\begin{figure}
\vspace{5mm}
\centerline{\includegraphics[width=8cm]{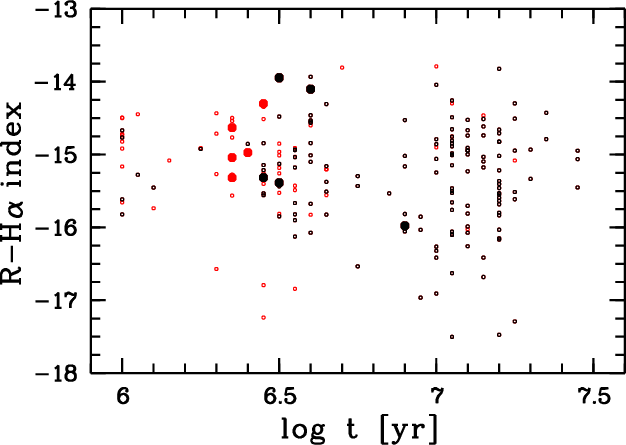}}
\caption{Relationship between the age of the stellar population in SFRs and the 
$R-{\rm H}\alpha$ index. Small circles are objects from the catalog of 
\citet{gusev2023}, large circles are objects in NGC~3963 and NGC~7292. Black denotes 
SFRs with an age estimation error $\Delta \log t<0.2$~dex, while red indicates 
those with an error greater than 0.2~dex.
\label{fig:rhat}}
\end{figure}

Figure~\ref{fig:rhat} shows the dependence between the age of an SFR and the 
$R-{\rm H}\alpha$ index. We used the $R-{\rm H}\alpha$ index instead of 
EW(H$\alpha$) due to greater uniformity in measurements of $R-{\rm H}\alpha$. 
The EW(H$\alpha$) value is sensitive to the choice of the background area 
subtracted from the spectrum continuum, and in the case of slit spectroscopy, 
it can also be affected by the inhomogeneity of the emission distribution in 
H$\alpha$ and the stellar continuum within the SFR. Criticism of using parameter 
EW(H$\alpha$) as an age indicator can be found in 
\citet{morisset2016,kreckel2022,scheuermann2023}.

Unlike the similar plot in \citet{gusev2023}, Fig.~\ref{fig:rhat} presents objects 
with any errors in the estimate of $t$; regions with $\Delta \log t>0.2$~dex are 
highlighted in red. Large circles denote 10 objects of class~2 with age estimates 
from galaxies investigated in this study. As seen from the figure, including new 
data does not change the conclusions drawn in \citet{gusev2023}: while the minimum 
value of the $R-{\rm H}\alpha$ index for SFRs of the corresponding age decreases 
with increasing $t$ (from $\approx-16$ for $t\approx2$~Myr to $\approx-17.5$ for 
$t\approx12$~Myr), the upper limit of $R-{\rm H}\alpha\approx-14$ (corresponding to 
EW(H$\alpha)\sim1000$\AA) does not depend on the age of the SFR. Thus, 
$R-{\rm H}\alpha$ index and EW(H$\alpha$) are poor indicators of the age of young 
stellar systems. The only conclusion possible from the analysis of Fig.~\ref{fig:rhat} 
is that the index $R-{\rm H}\alpha<-16$ (EW(H$\alpha)<100$\AA) indicates an age of 
the SFRs greater than 3~Myr.

\begin{figure}
\vspace{5mm}
\centerline{\includegraphics[width=8cm]{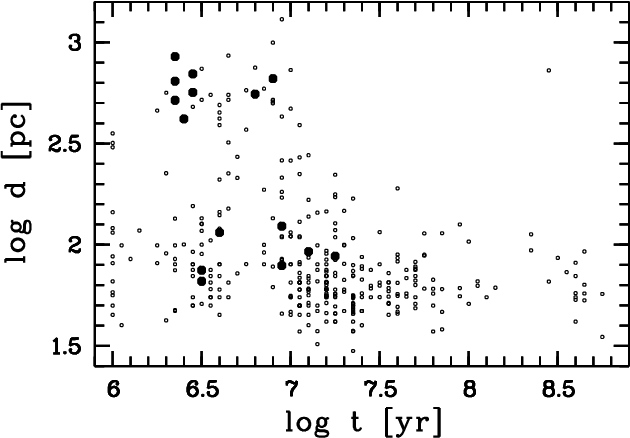}}
\caption{Size--age relationship for SFRs from the complete catalog sample 
(small circles) and regions in NGC~3963 and NGC~7292 (large black circles). 
Multi-component SFRs (e.g., double, triple) are not shown on the plot.
\label{fig:dt}}
\end{figure}

Among the SFRs in our sample, there are both large star complexes and aggregates 
-- compound systems including groups of OB associations and young clusters -- and 
individual star associations and clusters. Gravitationally unbound star associations 
expand with age \citep{efremov1998,zwart2010}. The sizes of young clusters are 
weakly dependent on age. The linear resolution of our observations, even in the 
nearest galaxies, is 30-40~pc and does not allow us to detect more compact star 
clusters. Differentiating the youngest stellar systems ($t<10$~Myr) by parameters 
into associations and clusters is difficult in principle \citep{gieles2011}, and 
with our linear resolution, it is impossible.

The position of the studied SFRs on the size--age diagram (Fig.~\ref{fig:dt}) 
illustrates the presence of various types of young stellar objects and their 
characteristics. A group of large star formation complexes ($d>400$~pc), including 
NGC~3963, stands out, and almost all of them are younger than 10~Myr. This is 
a consequence of two main factors. First, star complexes, with very rare 
exceptions, are relatively short-lived formations, disintegrating within 
100-200~Myr \citep{efremov1995}. Second, star formation in them does not occur 
simultaneously. The photometric age estimate of a complex is mainly influenced 
by the most recent star formation bursts in its components. Selection effects must 
also be considered: most SFRs were identified based on emission in the H$\alpha$ line.

Most SFRs with diameters up to 100~pc and older than 10~Myr are apparently young 
star clusters. In Fig.~\ref{fig:dt}, no dependency between age and size is observed 
for them. Note the absence of compact objects with $d<80$~pc among SFRs aged 4-8~Myr. 
HII~regions of this age are in the phase of expanding ionized gas envelopes, whose 
emission contributes to the total flux in photometric bands $U$, $B$, and others. The 
sizes of SFRs at this evolutionary stage, as determined from our images in the band, 
are effectively the sizes of the expanding gas envelopes.

The mass--size dependences for SFRs and their progenitors -- giant molecular clouds 
(GMCs) -- also have significant fundamental importance. The correlation $M\sim d^2$ 
between the masses and sizes of GMCs has been known since \citet{larson1981}. A close 
relationship between $M$ and $d$ was also obtained for SFRs: $M\sim d^{2.33\pm0.19}$ 
\citep{bastian2005}, $M\sim d^{2.0\pm0.3}$ \citep{adamo2013}. Although recent studies 
show more complex relationships between masses and sizes in both gas and stellar 
conglomerates \citep[see review in][]{grudic2021}, the dependence $M\sim d^2$ for 
GMCs and SFRs is well explained within modern interstellar medium physics theories. 
The shift of $\sim1.5$~dex between the relationships for GMCs and SFRs is explained by 
the inefficiency of star formation ($<10\%$) in GMCs 
\citep{larson1981,bastian2005,adamo2013}.

\begin{figure}
\vspace{5mm}
\centerline{\includegraphics[width=8cm]{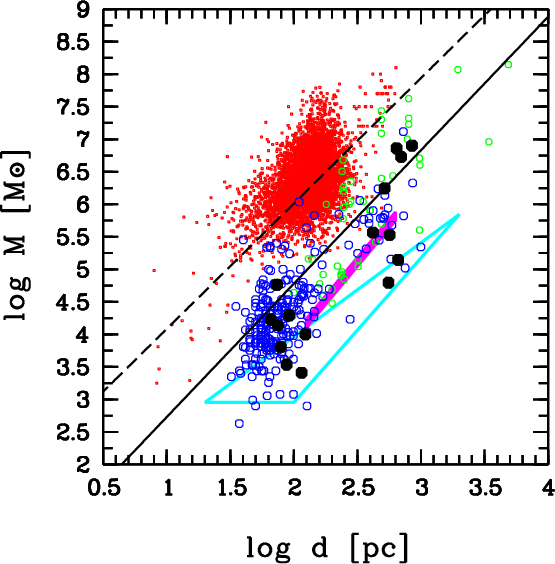}}
\caption{Mass--size relationship for GMCs 
\citep[red dots from][]{bolatto2008,wei2012,rosolowsky2021} and SFRs (green circles 
\citep{adamo2013}, purple parallelogram \citep{bastian2005}, region within the cyan 
triangle from \citet{gouliermis2017}). Blue circles represent single SFRs from our 
catalog \citep{gusev2023} with $\Delta M/M<0.2$; black circles represent SFRs in 
NGC~3963 and NGC~7292. The solid black line represents the $M\sim d^{2.0}$ 
relationship for young massive star clusters according to \citet{adamo2013}; the 
dashed black line represents the $M\sim d^{1.9}$ relationship for GMCs according to 
\cite{bolatto2008,adamo2013}.
\label{fig:dm}}
\end{figure}

Figure~\ref{fig:dm} shows the mass--size diagram for numerous samples of GMCs 
\citep{bolatto2008,wei2012,rosolowsky2021} and HII~regions 
\citep{bastian2005,adamo2013,gouliermis2017}. Unlike Fig.~20 in \citet{gusev2023}, 
where we included only isolated single SFRs from our catalog, here we also added 
single objects -- parts of larger SFRs. The mass--size plot shows that the positions 
of SFRs in our sample (blue and black circles) best match the results of 
\citet{adamo2013} (green circles), extending their dependence for complexes toward 
smaller star associations. Our results for some catalog objects are not inconsistent 
with data from \citet{bastian2005} (purple parallelogram) and \citet{gouliermis2017} 
(cyan triangle). However, effects related to the linear resolution of our 
observations play a significant role here: for SFRs with diameters smaller than the 
resolution limit, the $d$ value is overestimated and equals the linear resolution 
for the corresponding galaxy. This effect is clearly visible in the case of SFRs from 
NGC~3963 and NGC~7292 (black circles in Fig.~\ref{fig:dm}), which do not show any 
dependence of $M$ on $d$ separately for SFR samples in distant NGC~3963 (upper right 
group of black circles in the figure) and nearby NGC~7292 (lower left group).

\section{CONCLUSIONS}

In this study, we present results from investigating the physical parameters of 
stellar populations in 93 SFRs of the large spiral galaxy NGC~3963 with signs of 
peculiarity and 64 SFRs in the Magellanic-type galaxy NGC~7292. The data obtained 
complement our research in \citet{gusev2023}, which studied 1510 SFRs in 19 galaxies, 
and are included in the third version of the catalog of photometric, physical, and 
chemical parameters of SFRs, available electronically on the website of the SAI MSU 
(http://lnfm1.sai.msu.ru/$\sim$gusev/sfr\_cat.html).

We focus on two key aspects of the methodology for estimating the age and mass of SFRs 
using stellar population evolutionary models: \\
(1) The use of the extinction measure in SFRs determined from the Balmer decrement is 
only accurate for HII~regions where the photometric centers of the gas emission (in the 
H$\alpha$ line) and the stars (in broad photometric bands) coincide. \\
(2) For analyzing stellar systems with masses less than $1\cdot10^4 M\sun$, it is 
necessary to use evolutionary models with a discrete (randomly populated) IMF.

Among the 157 identified SFRs in the galaxies, we were able to obtain metallicity 
estimates for 27, extinction measurements from the Balmer decrement for 33, mass 
estimates for 16, and age estimates for 15 SFRs.

The main findings of the study are as follows. \\
(1) The number of HII~regions with coinciding and shifted centers of gas and stellar 
emission is approximately equal. This likely reflects the approximate equality in the 
time that a young stellar region spends at corresponding evolutionary stages. \\
(2) The diameter of SFRs in NGC~3963, $650\pm200$~pc, is typical for star complexes, 
while in NGC~7292, $75\pm20$~pc, it is typical for star associations. \\
(3) The stellar population masses in the studied SFRs in the galaxy NGC~3963 range 
from $6\cdot10^4 M\sun$ to $8\cdot10^6 M\sun$, while in the SFRs of NGC~7292 they 
range from $3\cdot10^3 M\sun$ to $6\cdot10^4 M\sun$. SFRs without emission in the 
H$\alpha$ line were, on average, less massive in both galaxies, which is likely 
explained by selection effects. \\
(4) The measured age of the stellar populations in the SFRs ranges from 2.2 to 18~Myr, 
with only two regions (both in NGC~7292) being older than 10~Myr. The age of SFRs 
is clearly correlated with the presence of emission in the H$\alpha$ line: HII~regions 
in galaxies younger than 6-8~Myr ($t=2.2-7.9$~Myr) show H$\alpha$ emission, while 
older regions ($t=6.3-18$~Myr) do not. \\
(5) We confirm the conclusion of \citet{gusev2023} that EW(H$\alpha$) and the 
$R-{\rm H}\alpha$ index are poor indicators of the age of a young stellar system. \\
(6) The positions of the SFRs in our sample on the mass--size diagram best match 
the results from \citet{adamo2013}, extending their dependence for complexes towards 
smaller star associations. Deviations from the dependence obtained in 
\citet{adamo2013} are caused by effects related to the linear resolution of our 
observations.

\section*{ACKNOWLEDGMENTS}

The authors thank the reviewer for valuable comments and suggestions. We also express 
our gratitude to A.~E.~Perepelitsin (SAO RAS) for technical support with observations 
using the MaNGaL instrument. This study used open data from the HyperLEDA 
(http://leda.univ-lyon1.fr) and NASA/IPAC Extragalactic Database 
(http://ned.ipac.caltech.edu), Padova group isochrones from the CMD server 
(http://stev.oapd.inaf.it), SExtractor software (http://sextractor.sourceforge.net), 
and the ESO-MIDAS image processing system (http://www.eso.org/sci/software/esomidas) 
developed at the Southern European Observatory.

\section*{FUNDING}

The development of the instrumentation base of the Caucasian Mountain Observatory 
of SAI MSU, is supported by the MSU Development Program. The processing and analysis 
of narrow-band images of NGC 3963 were carried out under the state assignment of the 
Special Astrophysical Observatory of the Russian Academy of Science, approved by the 
Ministry of Science and Higher Education of the Russian Federation. VSK thanks the 
Fund for the Development of Theoretical Physics and Mathematics ''BAZIS'' (project 
no.~23-2-2-6-1) for support.

\section*{CONFLICT OF INTEREST}

The authors of this work declare that they have no conflicts of interest.

\vspace{4mm}
\hspace{3.5cm}
{\it Translated by M.~Chubarova}

\vspace{4mm}
{\bf Publisher's Note.} Pleiades Publishing remains neutral with regard 
to jurisdictional claims in published maps and institutional affiliations.

\end{document}